\documentclass[doublecol,linenumbers]{epl2} 

\title{Finite-velocity diffusion on a comb}
\shorttitle{Finite-velocity diffusion on a comb} 

\author{T. Sandev\inst{1,2,3} \and A. Iomin\inst{4}}
\shortauthor{T. Sandev and A. Iomin}

\institute{                    
  \inst{1} Radiation Safety
  Directorate, Partizanski odredi 143, P.O. Box 22, 1020 Skopje,
  Macedonia\\
  \inst{2}Institute of Physics, Faculty of Natural Sciences and Mathematics, Ss Cyril and Methodius University, Arhimedova 3, 1000 Skopje, Macedonia\\
  \inst{3}Research Center for Computer Science and Information Technologies, Macedonian Academy of Sciences and Arts, Bul. Krste Misirkov 2, 1000 Skopje,
  Macedonia\\
  \inst{4}Department of Physics, Technion, Haifa 32000, Israel
}

\abstract{
A Cattaneo equation for a comb structure is considered. We present a rigorous analysis of the obtained fractional diffusion equation, and corresponding solutions for the probability distribution function are obtained in the form of the Fox $H$-function and its infinite series. The mean square displacement along the backbone is obtained as well in terms of the infinite series of the Fox $H$-function. The obtained solutions describe the transition from normal diffusion to subdiffusion, which results from the comb geometry.}

\begin{document}

\maketitle

\section{Introduction}

It is well known that any compact initial condition, evolving due to a parabolic (diffusion) equation, ``spreads'' instantly  to infinity. This situation relates for example to the Fickian law of diffusion, or Fourier's law of heat conduction, according equation $J=-\mathcal{D}\nabla P$, where $\mathcal{D}$ is the diffusion coefficient, $P=P(x,t)$ is the probability distribution function (PDF), while $J=J(x,t)$ is the probability flux. To overcome this unrealistic singularity with the infinite velocity of propagation, the so called telegrapher's or Cattaneo equation has been introduced \cite{cattaneo1,cattaneo2}
\begin{eqnarray}\label{classical cattaneo eq}
\frac{\partial}{\partial t}u(x,t)+\tau\frac{\partial^{2}}{\partial t^{2}}u(x,t)=\mathcal{D}\frac{\partial^{2}}{\partial x^{2}}u(x,t),
\end{eqnarray}
where $\tau$ is a characteristic time constant, $\mathcal{D}$ is the diffusion coefficient, which relates to a finite propagation velocity $v=\sqrt{\frac{\mathcal{D}}{\tau}}$. This equation is considered for  various realizations of initial and boundary conditions. In the diffusion limit ($\tau\rightarrow0$) of infinite-velocity propagation one recovers the standard diffusion equation. In the opposite limit $\tau\rightarrow\infty$, it becomes the wave equation with $\mathcal{D}/\tau=v^2$ being finite squared speed of the wave. The telegrapher’s equation was proposed by Kelvin and of Heaviside in electrodynamics theory and it was essentially employed in the heat transfer theory \cite{ref1} and persistent random walk \cite{cattaneo1,cattaneo2,ref1}. The standard Cattaneo equation has been solved analytically \cite{masoliver_weiss,weiss}, and further generalization on fractional order equations has been done by several authors \cite{masoliver_lindenberg,metzler_compte,qi1,qi2,qi4,flux1,zorica,flux2,ramirez,colin,siam}. 

Our main concern of the Cattaneo Eq.~(\ref{classical cattaneo eq}) in the comb geometry is motivated by ionic transport inside neuron dendrite structure. Recent experiments, together with numerical simulations, have investigated the calcium transport inside spiny dendrites to understand the role of calcium in signal transmission and neural plasticity. This issue is well reviewed in Refs. \cite{segal,segal2}. Based on these experimental finding, different theoretical approaches have been developed to explore the transport properties of spiny dendrites. It is an active field of study \cite{iomin_csf,iomin_pre2013,yuste,rose}, and new experimental findings on calcium transport and reaction transport in neuroscience \cite{rose,bresslof} pose new questions to understand the impact of the geometry on calcium transport and reactions in spiny dendrites and the extension of various reaction-transport models to the case of subdiffusion. Recent experiments established a relation between the geometry of the dendrite spines and the subdiffusion observed in Refs. \cite{santamaria,santamaria2}. It supports the application of the comb model, which is a convenient tool to explore analytically the anomalous transport in spiny dendrites in the framework of the continuous time random walk (CTRW) approach. Comb-like models can mimic ramified structures such as spiny dendrites of neuron cells \cite{iomin_csf,iomin_pre2013,yuste} and can be used to describe the movement and binding dynamics of particles inside the dendritic spines. The corresponding process along the backbone is subdiffusiove with the power law evolution of the mean square displacement $$\left\langle x^{2}(t)\right\rangle\simeq t^{1/2}.$$

In this paper we consider finite-velocity diffusion on a comb, described by a comb model
\begin{eqnarray}\label{cattaneo eq comb} 
\frac{\partial}{\partial t}P(x,y,t)+\tau\frac{\partial^{2}}{\partial t^{2}}P(x,y,t)&=&\mathcal{D}_{x}\delta(y)\frac{\partial^{2}}{\partial x^{2}}P(x,y,t)\nonumber\\&+&\mathcal{D}_{y}\frac{\partial^{2}}{\partial y^{2}}P(x,y,t),
\end{eqnarray}
which represents the standard Cattaneo equation on a comb structure. The $\delta$-function means that the diffusion along the $x$-direction is allowed only at $y=0$ (the backbone). The particle moves in the backbone and can eventually be trapped in the fingers. The initial condition is given by
\begin{equation}
\label{initial condition}
P(x,y,t=0)=\delta(x)\delta(y), \quad \frac{\partial}{\partial t}P(x,y,t=0)=0,
\end{equation}
and the boundary conditions for the PDF $P(x,y,t)$ and for $\frac{\partial}{\partial q}P(x,y,t)$, $q=\{x,y\}$ are set to zero at infinities, $x=\pm\infty$, $y=\pm\infty$. The diffusion coefficient along the $x$-direction is $\mathcal{D}_{x}\delta(y)$, with physical dimension $[\mathcal{D}_{x}\delta(y)]=\mathrm{m}^{2}/\mathrm{s}$, i.e., $[\mathcal{D}_{x}]=\mathrm{m}^{3}/\mathrm{s}$ ($[\delta(y)]=\mathrm{m}^{-1}$). The diffusion coefficient along the fingers is $\mathcal{D}_{y}$, $[\mathcal{D}_{y}]=\mathrm{m}^{2}/\mathrm{s}$. Correspondingly, the finite propagation velocities are $v_{x}=\mathcal{D}_x/[\tau\mathcal{D}_y]= \mathcal{D}_x/[v_y^2\tau^2]$ and $v_{y}=\sqrt{\mathcal{D}_{y}/\tau}$, where the $x$ component of the velocity accounts also the comb geometry, when transport along the backbone depends also on transport in fingers. Note also that the relation between the real three dimensional Laplace operator and the Laplace operator in the comb model (2)  was established  in Ref.~\cite{iomin_new}.

\section{Cattaneo equation}

Prior concerning with the comb model~(\ref{cattaneo eq comb}) we discuss the properties of the telegrapher's equation (\ref{classical cattaneo eq}) with the initial conditions of Eq.~(\ref{initial condition}) and  natural (zero) boundary conditions at infinity. This equation was considered in Ref. [4], where an exact solution was obtained in the framework of Bessel functions. Here we give the solution in terms of the Fox $H$-function defined by inverse Mellin transform \cite{saxena_book}:
\begin{eqnarray}\label{H_integral}
H_{p,q}^{m,n}(z)&=&H_{p,q}^{m,n}\left[z\left|\begin{array}{c l}
    (a_1,A_1), (a_2,A_2),\dots,(a_p,A_p)\\
    (b_1,B_1), (b_2,B_2),\dots,(b_q,B_q)
  \end{array}\right.\right]\nonumber\\&=&\frac{1}{2\pi\imath}\int_{\Omega}\theta(s)z^{-s}\,ds,
\end{eqnarray}
where
\begin{eqnarray}\label{theta}
\theta(s)=\frac{\prod_{j=1}^{m}\Gamma(b_j+B_js)\prod_{j=1}^{n}\Gamma(1-a_j-A_js)}{\prod_{j=m+1}^{q}\Gamma(1-b_j-B_js)\prod_{j=n+1}^{p}\Gamma(a_j+A_js)},
\end{eqnarray}
and conditions of the integration are fulfilled\footnote{The conditions are $0\leq n\leq p$, $1\leq m\leq q$, $a_i,b_j \in \mathrm{C}$, $A_i,B_j\in\mathrm{R}^{+}$, $i=1,...,p$, $j=1,...,q$. Contour integration $\Omega$ starts at $c-\imath\infty$ and finishes at $c+\imath\infty$ separating the poles of the function $\Gamma(b_j+B_js)$, $j=1,...,m$ with those of the function $\Gamma(1-a_i-A_is)$, $i=1,...,n$.}. 

After Fourier-Laplace transforms, one obtains the solution of Eq.~(\ref{classical cattaneo eq}) in the $(k,s)$-space
\begin{equation}\label{cattaneo eq FL}
\tilde{\hat{P}}(k,s)=\frac{1+s\tau}{s(1+s\tau)+\mathcal{D}k^{2}}.
\end{equation}
Its inverse Fourier transform yields
\begin{eqnarray}\label{cattaneo eq L}
\hat{P}(x,s)&=&\frac{1}{2v}\left[1+(s\tau)^{-1}\right]^{1/2}\nonumber\\&\times &\exp\left(-\frac{s\left[1+(s\tau)^{-1}\right]^{1/2}}{v}|x|\right).
\end{eqnarray}
Using the variable change $z=1+(s\tau)^{-1}$ and $\rho=\frac{|x|}{v}s$, we present Eq.~(\ref{cattaneo eq L}) in the compact form
\begin{eqnarray}\label{cattaneo eq L2}
\hat{P}(x,s)=\frac{1}{2v}\sqrt{z}e^{-\rho\sqrt{z}}.
\end{eqnarray}
Now we apply Mellin transform to Eq.~(\ref{cattaneo eq L2}). Using a trick, we have 
\begin{eqnarray}
&&\hat{P}(x,s)=\mathcal{M}^{-1}\left[\mathcal{M}[\hat{P}(x,s)](\xi)\right](s)\nonumber\\&&=\frac{1}{2v}\mathcal{M}^{-1}\left[\int_0^{\infty}z^{1/2+\xi-1}e^{-\rho\sqrt{z}}\,dz\right](s)\nonumber\\&&=\mathcal{M}^{-1}\left[\frac{\rho^{-2\xi-1}}{v}\Gamma(2\xi+2)\right](s)\,.
\end{eqnarray}
This eventually yields the PDF in the Laplace space in the form of the 
Fox $H$-function
\begin{eqnarray}\label{cattaneo eq L3}
\hat{P}(x,s)&=&\frac{1}{2\pi i}\frac{1}{\rho v}\int_{\Omega}
\Gamma(2\xi+2)(\rho^{2})^{-\xi}\,d\xi\nonumber\\  &=&\frac{1}{v\rho}H_{0,1}^{1,0}\left[\rho^{2}z\left|\begin{array}{c l}
    -\\
    (1,2)
  \end{array}\right.\right].
\end{eqnarray}
Here we used the definition of the Fox $H$-function in Eq. (\ref{H_integral}). Expanding function  $\hat{P}(x,s)$ in Taylor series about $z=1$, we have
\begin{eqnarray}\label{cattaneo eq L Taylor}
&&\hat{P}(x,s)=\frac{1}{v\rho}\sum_{k=0}^{\infty}\frac{(z-1)^{k}}{k!}\nonumber\\&&\times\frac{d^{k}}{dz^{k}}\left\{H_{0,1}^{1,0}\left[\rho^{2}z\left|\begin{array}{c l}
    -\\
    (1,2)
  \end{array}\right.\right]\right\}_{z=1}\nonumber\\&&=\frac{1}{|x|}\sum_{k=0}^{\infty}\frac{1}{k!}\tau^{-k}s^{-k-1}H_{1,2}^{1,1}\left[\frac{x^{2}}{v^{2}}s^{2}\left|\begin{array}{c l}
                  (0,1)\\
                  (1,2), (k,1)
                \end{array}\right.\right].\nonumber\\
\end{eqnarray}
Application of the inverse Laplace transformation in Eq.~(\ref{cattaneo eq L Taylor}) yields
\begin{eqnarray}\label{cattaneo eq final solution}
P(x,t) &=&\frac{1}{2|x|}\sum_{k=0}^{\infty}\frac{(-1)^{k}}{k!}\left(\frac{t}{\tau}\right)^{k}\nonumber\\&\times & H_{2,2}^{2,0}\left[\frac{|x|}{vt}\left|
\begin{array}{c l}
         (k+1,1), (0,1/2)\\
         (k,1/2), (1,1)
\end{array}\right.\right].
\end{eqnarray}

The asymptotic behavior of the solution in the case of $t/\tau\rightarrow\infty$ (i.e., $s\tau\rightarrow0$) becomes Gaussian
\begin{eqnarray}\label{cattaneo eq L large t}
P(x,t)\simeq\frac{1}{\sqrt{4\pi v^{2}\tau t}}\exp\left(-\frac{x^{2}}{4v^{2}\tau t}\right),
\end{eqnarray}
while the opposite case of $t/\tau\rightarrow0$ (i.e., $s\tau\rightarrow\infty$), from Eq.~(\ref{cattaneo eq final solution}), yields
\begin{eqnarray}\label{cattaneo eq L small t}
P(x,t)&\simeq&\frac{1}{|x|}H_{0,0}^{0,0}\left[\frac{|x|^2}{(vt)^2}\left|
\begin{array}{c l}
         -\\
         -
\end{array}\right.\right]=\frac{\delta\left(1-\frac{|x|^2}{(vt)^2}\right)}{|x|}\nonumber\\&=&\frac{1}{2}\left[\delta(x+vt)+\delta(x-vt)\right],
\end{eqnarray}
where we use the result given in \cite{fcaa_delta} and properties of the Dirac delta function.

The MSD can be expressed in the form of the two parameter Mittag-Leffler function, as well. The MSD reads $$\left\langle x^{2}(t)\right\rangle=\mathcal{L}^{-1}\left.\left[-\frac{\partial^{2}}{\partial k^{2}}\tilde{\hat{P}}(k,s)\right]\right|_{k=0},$$ from where we find
\begin{eqnarray}
\label{msd cattaneo eq}
\left\langle x^{2}(t)\right\rangle&=&\frac{2\mathcal{D}}{\tau}\mathcal{L}^{-1}\left[\frac{s^{-1}}{s+\tau^{-1}}\right]\nonumber\\&=&2\mathcal{D}\tau\left(\frac{t}{\tau}\right)^{2}E_{1,3}\left(-\frac{t}{\tau}\right)\nonumber\\&=&2\mathcal{D}\left[t+\tau\left(e^{-t/\tau}-1\right)\right].
\end{eqnarray}   
Here $E_{\alpha,\beta}(z)$ is the two parameter Mittag-Leffler function, defined by \cite{hilfer}
\begin{equation}\label{two ml}
E_{\alpha,\beta}(z)=\sum_{k=0}^{\infty}\frac{z^k}{\Gamma(\alpha k+\beta)},
\end{equation}
where $z, \beta \in \mathrm{C}$, and $\Re(\alpha)>0$. Note that the two parameter Mittag-Leffler function (\ref{two ml}) is a generalization of the one parameter Mittag-Leffler function $E_{\alpha}(z)=E_{\alpha,1}(z)$ and the exponential function $E_{1,1}(z)=e^{z}$.

For the short time limit $t/\tau\rightarrow0$ the MSD corresponds to ballistic motion $$\left\langle x^{2}(t)\right\rangle\simeq\mathcal{D}\tau\left(\frac{t}{\tau}\right)^{2},$$ and then it changes to normal diffusion $$\left\langle x^{2}(t)\right\rangle\simeq2\mathcal{D}t$$ in the long time limit  $t/\tau\rightarrow\infty$. These asymptotic diffusion and wave limits coincide exactly with results obtained in Ref.~[4]. Here we note that the same result we obtain if one applies the exact solution (\ref{cattaneo eq final solution}), that is
\begin{eqnarray}\label{cattaneo eq msd from final}
\left\langle x^{2}(t)\right\rangle &=&\int_{-\infty}^{\infty}x^{2}P(x,t)\,dx=2(vt)^{2}\sum_{k=0}^{\infty}\frac{(-t/\tau)^{k}}{\Gamma(k+3)}\nonumber\\&=&2\mathcal{D}\tau\left(\frac{t}{\tau}\right)^{2}E_{1,3}\left(-\frac{t}{\tau}\right),
\end{eqnarray}
where we use the Mellin transform of the Fox $H$-function, in definition~(\ref{theta}).

\section{Solution of Cattaneo equation for a comb structure}

Let us now analyze the comb Cattaneo equation (\ref{cattaneo eq comb}) with the initial conditions (\ref{initial condition}) and zero boundary conditions at infinity. Fourier-Laplace transforming we find
\begin{eqnarray}
\label{cattaneo eq comb FFL}
\bar{\tilde{\hat{P}}}(k_x,k_y,s)=\frac{1+s\tau-\mathcal{D}_{x}k_{x}^{2}\tilde{\hat{P}}(k_x,y=0,s)}{s(1+s\tau)+\mathcal{D}_{y}k_{y}^{2}},
\end{eqnarray}
which yields after inverse Fourier transform in respect to $k_y$
\begin{eqnarray}
\label{cattaneo eq comb FL}
&&\tilde{\hat{P}}(k_x,y,s)=\frac{1}{2\sqrt{\mathcal{D}_{y}}}\frac{\left[1+s\tau-\mathcal{D}_{x}k_{x}^{2}\tilde{\hat{P}}(k_x,y=0,s)\right]}{\sqrt{s(1+s\tau)}}\nonumber\\&&\times\exp\left(-\frac{\sqrt{s(1+s\tau)}}{\sqrt{\mathcal{D}_{y}}}|y|\right).
\end{eqnarray}
At this first step of finding a closed form of the solution, we need to find $\hat{P}(x,y=0,s)$. From Eq.~(\ref{cattaneo eq comb FL}) we have
\begin{eqnarray}
\label{cattaneo eq comb FL2}
\tilde{\hat{P}}(k_x,y=0,s)=\frac{1}{2\sqrt{\mathcal{D}_{y}}}\frac{1+s\tau}{\sqrt{s(1+s\tau)}+\frac{\mathcal{D}_{x}}{2\sqrt{\mathcal{D}_{y}}}k_{x}^{2}},
\end{eqnarray}
which eventually yields the closed form of the PDF in the Fourier-Laplace space
\begin{eqnarray}
\label{cattaneo eq comb FL3}
\bar{\tilde{\hat{P}}}(k_x,k_y,s)&=&\frac{1+s\tau}{s(1+s\tau)+\mathcal{D}_{y}k_{y}^{2}}\nonumber\\&\times&\frac{\sqrt{s(1+s\tau)}}{\sqrt{s(1+s\tau)}+\frac{\mathcal{D}_{x}}{2\sqrt{\mathcal{D}_{y}}}k_{x}^{2}}.
\end{eqnarray}
Integrating Eq.~(\ref{cattaneo eq comb}) in respect to $y$ we obtain the marginal PDF $$p_{1}(x,t)=\int_{-\infty}^{\infty}P(x,y,t)\,dy.$$ Its Fourier-Laplace image results from Eq.~(\ref{cattaneo eq comb FL3}) as follows\footnote{It is also known as the Montroll-Weiss equation \cite{schneider1}.}
\begin{equation}
\label{p1 FL}
\tilde{\hat{p}}_{1}(k_x,s)=\bar{\tilde{\hat{P}}}(k_x,k_y=0,s)=\frac{1}{s}\frac{\sqrt{s(1+s\tau)}}{\sqrt{s(1+s\tau)}+\frac{\mathcal{D}_{x}}{2\sqrt{\mathcal{D}_{y}}}k_{x}^{2}},
\end{equation}
and correspondingly the Laplace image is
\begin{eqnarray}
\label{p1 L}
\hat{p}_{1}(x,s) &=&\frac{1}{2}\sqrt{\frac{2\sqrt{\mathcal{D}_{y}\tau}}{\mathcal{D}_{x}}}s^{-1/2}[1+(s\tau)^{-1}]^{1/4}\nonumber\\&\times&\exp\left(-\sqrt{\frac{2\sqrt{\mathcal{D}_{y}\tau}}{\mathcal{D}_{x}}}s^{1/2}[1+(s\tau)^{-1}]^{1/4}|x|\right).\nonumber\\
\end{eqnarray}
The inverse Laplace transform can be found in the same way as it was done for the classical Cattaneo equation, cf. Eqs.~(\ref{cattaneo eq L2})-(\ref{cattaneo eq L Taylor}). Rewriting Eq.~(\ref{p1 L}) in the compact form by means of the variable change $z=1+(s\tau)^{-1}$ and $\rho=\sqrt{\frac{2\sqrt{\mathcal{D}_{y}\tau}}{\mathcal{D}_{x}}}|x|s^{1/2}$, we obtain $\hat{p}_1(x,s)$ in the form of the Fox H-function
\begin{eqnarray}\label{cattaneo eq L2 comb}
\hat{p}_{1}(x,s)=\frac{\rho\,z^{1/4}}{2s|x|}e^{-\rho\,z^{1/4}}=\frac{2}{s|x|}H_{0,1}^{1,0}\left[\rho^{4}z\left|\begin{array}{c l}
    -\\
    (1,4)
  \end{array}\right.\right].\nonumber\\
\end{eqnarray}
Its Taylor expansion about $z=1$ gives
\begin{eqnarray}\label{cattaneo eq L Taylor comb}
\hat{p}_{1}(x,s) &=&\frac{2}{|x|}\sum_{k=0}^{\infty}\frac{\tau^{-k}}{k!}s^{-k-1}\nonumber\\&\times&H_{1,2}^{1,1}\left[\left(\frac{2\sqrt{\mathcal{D}_{y}\tau}}{\mathcal{D}_{x}}\right)^{2}|x|^{4}s^{2}\left|\begin{array}{c l}
          (0,1)\\
          (1,4), (k,1)
        \end{array}\right.\right].\nonumber\\
        \end{eqnarray}
Now inverse Laplace transform yields the solution for the marginal PDF
\begin{eqnarray}\label{cattaneo eq final solution comb}
p_{1}(x,t) &=&\frac{1}{|x|}\sum_{k=0}^{\infty}\frac{(-1)^{k}}{k!}\left(\frac{t}{\tau}\right)^{k}\nonumber\\&\times&H_{2,2}^{2,0}\left[\frac{2\sqrt{\mathcal{D}_{y}\tau}}{\mathcal{D}_{x}}\frac{x^{2}}{t}\left|\begin{array}{c l}
                                  (k+1,1), (0,1/2)\\
                                  (k,1/2), (1,2)
                                \end{array}\right.\right].\nonumber\\
\end{eqnarray}

The asymptotic behavior of the PDF for $t/\tau\rightarrow\infty$ ($s\tau\rightarrow0$) reads
\begin{eqnarray}
\label{p1 L large t}
p_{1}(x,t) \simeq\frac{1}{2|x|}H_{1,1}^{1,0}\left[\frac{1}{\sqrt{\frac{\mathcal{D}_{x}}{2\sqrt{D}_{y}}}}\frac{|x|}{t^{1/4}}\left|\begin{array}{c l}
                                  (1,1/4)\\
                                  (1,1)
                                \end{array}\right.\right],
\end{eqnarray}
which is exactly the PDF for a diffusion equation on a comb \cite{mmnp}. Correspondingly, the short time solution for $t/\tau\rightarrow0$ ($s\tau\rightarrow\infty$) becomes
\begin{eqnarray}
\label{p1 L small t}
p_{1}(x,t) \simeq\frac{1}{\sqrt{4\pi\frac{\mathcal{D}_{x}}{2\sqrt{\mathcal{D}_{y}\tau}}t}}\exp\left(-\frac{x^{2}}{4\frac{\mathcal{D}_{x}}{2\sqrt{\mathcal{D}_{y}\tau}}t}\right).
\end{eqnarray}
Comparing these limiting cases with corresponding limiting results in Eqs.~(13) and (14), we conclude that the comb geometry affects strongly this wave-diffusion process. The wave dynamics is attenuated and fractional dynamics becomes dominant. It is also in a good qualitative agreement with experimental data on neocortical pyramidal neurons adapting with a time scale\footnote{We do not discuss this issue in the paper since it deserves a separate study.}, which is consistent with fractional order differentiation, such that the neuron's firing rate is a fractional derivative of slowly varying stimulus parameters \cite{new}.

The fractional Cattaneo equation can be obtained as follows. Let us rewrite the Montroll-Weiss equation~(\ref{p1 FL}) as follows
\begin{equation}
\label{p1 FL 2}
[1+(s\tau)^{-1}]^{1/2}\left[s\tilde{\hat{p}}_{1}(k_x,s)-1\right]=-\frac{\mathcal{D}_{x}}{2\sqrt{\mathcal{D}_{y}}}\sqrt{\tau}k_{x}^{2}\tilde{\hat{p}}_{1}(k_x,s),
\end{equation}
from where by inverse Fourier-Laplace transform we find the following generalized diffsuion equation
\begin{equation}
\label{p1 eq}
\int_{0}^{t}\gamma(t-t')\frac{\partial}{\partial t'}p_{1}(x,t')\,dt'=\frac{\mathcal{D}_{x}}{2\sqrt{\mathcal{D}_{y}}}\sqrt{\tau}\frac{\partial^{2}}{\partial x^{2}}p_{1}(x,t),
\end{equation}
where \begin{eqnarray}
\gamma(t)=\mathcal{L}^{-1}\left[\frac{s^{-1/2}}{(s+\tau^{-1})^{-1/2}}\right]=\left(\frac{t}{\tau}\right)^{-1}E_{1,0}^{-1/2}\left(-\frac{t}{\tau}\right),
\end{eqnarray}
while $E_{\alpha,\beta}^{\delta}(z)$ is the three parameter Mittag-Leffler function \cite{prabhakar}
\begin{equation}\label{three ml}
E_{\alpha,\beta}^{\delta}(z)=\sum_{k=0}^{\infty}\frac{(\delta)_k}{\Gamma(\alpha k+\beta)}\frac{z^k}{k!}.
\end{equation}
Here $\beta, \delta, z \in \mathrm{C}$, $\Re(\alpha)>0$, $(\delta)_{k}$ is the Pochhammer symbol $(\delta)_{0}=1$, $(\delta)_{k}=\frac{\Gamma(\delta+k)}{\Gamma(\delta)}$. Laplace transform of the Mittag-Leffler function reads
\begin{eqnarray}\label{Laplace ML3_1}
\mathcal{L}\left[t^{\beta-1}E_{\alpha,\beta}^{\delta}\left(-\lambda{t}^{\alpha}\right)\right]=\frac{s^{\alpha\delta-\beta}}{(s^\alpha+\lambda)^\delta}, \quad |\lambda/s^{\alpha}|<1.\nonumber\\
\end{eqnarray}
Note also that $E_{\alpha,\beta}^{1}(z)=E_{\alpha,\beta}(z)$, and it is a special case of the Fox $H$-function \cite{saxena_book}
\begin{equation}\label{ml H}
E_{\alpha,\beta}^{\delta}\left(-z\right)=\frac{1}{\delta}H_{1,2}^{1,1}\left[z\left|\begin{array}{l}
    (1-\delta,1)\\
    (0,1),(1-\beta,\alpha)
  \end{array}\right.\right].
\end{equation}

Eq.~(\ref{p1 eq}) can be rewritten as follows
\begin{equation}
\label{p1 eq Pr}
{_C}\mathcal{D}_{1,-\tau^{-1},0+}^{1/2,1}p_{1}(x,t)=\frac{\mathcal{D}_{x}}{2\sqrt{\mathcal{D}_{y}\tau}}\frac{\partial^{2}}{\partial x^{2}}p_{1}(x,t),
\end{equation}
where
\begin{eqnarray}\label{prabhakar derivative m=1}
&&{_{C}}\mathcal{D}_{\rho,-\nu,0+}^{\delta,\mu}f(t)\nonumber\\&&=\int_{0}^{t}(t-t')^{-\mu}E_{\rho,1-\mu}^{-\delta}\left(-\nu(t-t')^{\rho}\right)\frac{d}{dt'}f(t')\,dt'\nonumber\\
\end{eqnarray}
is the regularized Prabhakar fractional derivative \cite{garra}, and its Laplace transform reads
\begin{eqnarray}\label{laplace prabhakar derivative m=1}
&&\mathcal{L}\left[{_{C}}\mathcal{D}_{\rho,-\nu,0+}^{\delta,\mu}f(t)\right]\nonumber\\&&=s^{-\rho\delta+\mu-1}\left(s^{\rho}+\nu\right)^{\delta}\left[s\hat{f}(s)-f(0+)\right].
\end{eqnarray}

This also results in the  normal diffusion equation in the short time dynamics $t/\tau\rightarrow0$, \begin{eqnarray}\label{p1 eq Pr short}
\frac{\partial}{\partial t}p_{1}(x,t)=\frac{\mathcal{D}_{x}}{2\sqrt{\mathcal{D}_{y}\tau}}\frac{\partial^{2}}{\partial x^{2}}p_{1}(x,t),
\end{eqnarray}
and  it corresponds to the time fractional  diffusion equation for $t/\tau\rightarrow\infty$,
\begin{equation}
\label{p1 eq Pr long}
{_C}\mathcal{D}_{0+}^{1/2}p_{1}(x,t)=\frac{\mathcal{D}_{x}}{2\sqrt{\mathcal{D}_{y}}}\frac{\partial^{2}}{\partial x^{2}}p_{1}(x,t),
\end{equation}
where ${_C}\mathcal{D}_{0+}^{1/2}$ is the Caputo fractional derivative \cite{hilfer}
\begin{eqnarray}\label{caputo m=1}
{_{C}}\mathcal{D}_{0+}^{\alpha}f(t)=\frac{1}{\Gamma(1-\alpha)}\int_{0}^{t}(t-t')^{-\alpha}\frac{d}{dt'}f(t')\,dt'.
\end{eqnarray}

The MSD can be also estimated rigorously. From Eq.~(\ref{p1 FL 2}), we have 
\begin{eqnarray}
\label{msd cattaneo eq comb}
\left\langle x^{2}(t)\right\rangle&=&\mathcal{L}^{-1}\left.\left[-\frac{\partial^{2}}{\partial k^{2}}\tilde{\hat{p}}_{1}(k,s)\right]\right|_{k=0}\nonumber\\&=&2\left(\frac{\mathcal{D}_{x}}{2\sqrt{\mathcal{D}_{y}}}\right)\mathcal{L}^{-1}\left[\frac{s^{-3/2}}{(s+\tau^{-1})^{1/2}}\right]\nonumber\\&=&2\left(\frac{\mathcal{D}_{x}}{2\sqrt{\mathcal{D}_{y}}}\sqrt{\tau}\right)\left(\frac{t}{\tau}\right)E_{1,2}^{1/2}\left(-\frac{t}{\tau}\right).
\end{eqnarray}
Taking into account the series representation of the three  parameter Mittag-Leffler function (\ref{three ml}) we find that for the short time limit $t/\tau\rightarrow0$ the MSD in Eq.~(\ref{msd cattaneo eq comb}) results in normal diffusion $$\left\langle x^{2}(t)\right\rangle\simeq2\left(\frac{\mathcal{D}_{x}}{2\sqrt{\mathcal{D}_{y}}}\sqrt{\tau}\right)\left(\frac{t}{\tau}\right),$$ and in the long time limit  $t/\tau\rightarrow\infty$ it changes to subdiffusion $$\left\langle x^{2}(t)\right\rangle\simeq2\left(\frac{\mathcal{D}_{x}}{2\sqrt{\mathcal{D}_{y}}}\sqrt{\tau}\right)\frac{(t/\tau)^{1/2}}{\Gamma(3/2)},$$ where we apply the asymptotic formula for the three parameter Mittag-Leffler function \cite{garrappa}
\begin{equation}\label{GML_formula}
E_{\alpha,\beta}^{\gamma}(-z)=\frac{z^{-\gamma}}{\Gamma(\gamma)}\sum_{n=0}^{\infty}\frac{\Gamma(\gamma+n)}{\Gamma(\beta-\alpha(\gamma+n))}\frac{(-z)^{-n}}{n!},
\end{equation} 
with $z>1$, and $0<\alpha<2$.

It is worth noting that the MSD corresponds to the Mellin transform 
of the exact solution (\ref{cattaneo eq final solution comb}), which relates to Eq.~(\ref{theta}) and yields
\begin{eqnarray}\label{cattaneo eq msd comb}
\left\langle x^{2}(t)\right\rangle &=&2\frac{\mathcal{D}_{x}}{2\sqrt{\mathcal{D}_{y}\tau}}t\sum_{k=0}^{\infty}\frac{(-t/\tau)^{k}}{k!}\frac{\Gamma(k+1/2)}{\Gamma(1/2)\Gamma(k+2)}\nonumber\\&=&2\left(\frac{\mathcal{D}_{x}}{2\sqrt{\mathcal{D}_{y}}}\sqrt{\tau}\right)\left(\frac{t}{\tau}\right)E_{1,2}^{1/2}\left(-\frac{t}{\tau}\right).
\end{eqnarray}

\section{Finite domain solution}

In reality, the dendrite structure has a finite length along the backbone. In this case, the finite boundary conditions affect strongly the transient diffusion-subdiffusion process described by Eq.~(\ref{p1 eq Pr}). One should also bear in mind that the spine-finger structure is finite as well. However, we suppose reasonably that the contaminant transport in this ramified structure is essentially slower than in the backbone, and the boundary conditions for the fingers are at $y=\pm\infty$. Therefore, now it is described by the marginal PDF in the framework of Eq.~(\ref{p1 eq Pr}) with initial condition $p_1(x,t=0)=\delta(x)$ in the range $-L<x<L$ with absorbing boundary conditions $p_1(x=\pm L,t)=0$. It means that once a transporting particle reaches a boundary, it is instantly removed from the boundary.

We use the method of separation of variables $p_{1}(x,t)=X(x)T(t)$. Therefore, we find
\begin{equation}
\label{p1 eq Pr finite domain}
\frac{{_C}\mathcal{D}_{1,-\tau^{-1},0+}^{1/2,1}T(t)}{T(t)}=\frac{2\sqrt{\mathcal{D}_{y}\tau}}{\mathcal{D}_{x}}\frac{X''(x)}{X(x)}=-\lambda,
\end{equation}
where $\lambda$ is a separation constant. From here we have a system of two equations
\begin{eqnarray}
&&{_C}\mathcal{D}_{1,-\tau^{-1},0+}^{1/2,1}T(t)+\lambda{T(t)}=0,\label{p1 eq Pr finite domain T}\\
&&X''(x)+\frac{2\sqrt{\mathcal{D}_{y}\tau}}{\mathcal{D}_{x}}\lambda{X(x)}=0.\label{p1 eq Pr finite domain X}
\end{eqnarray}
Equation (\ref{p1 eq Pr finite domain X}) is the eigenvalue problem with the boundary condition $X(x=\pm L)=0$, which yields a set of eigenfunctions $X_n$ with corresponding eigenvalues $\lambda_n$. Thus, the solution of Eq.~(\ref{p1 eq Pr}) in the finite domain is $$p_{1}(x,t)=\sum_{n=0}^{\infty}T_{n}(t)X_{n}(x),$$ where $T_n(t=0)=1\, , \forall n$. Eventually, after accounting for the initial condition $p_{1}(x,0)=\delta(x)$, the solution reads
\begin{eqnarray}\label{solution_1}
p_{1}(x,t)=\frac{1}{L}\sum_{n=-\infty}^{\infty}e^{\imath\frac{(2n+1)\pi x}{2L}}T_{n}(t).
\end{eqnarray}
The solution of Eq.~(\ref{p1 eq Pr finite domain T}) can be found by the Laplace transform method. Thus, from relation (\ref{laplace prabhakar derivative m=1}), we have
\begin{eqnarray}
\left[1+(s\tau)^{-1}\right]^{1/2}\left[s\hat{T}_{n}(s)-1\right]+\lambda_{n}\hat{T}_{n}(s)=0,
\end{eqnarray}
i.e.,
\begin{eqnarray}
\hat{T}_{n}(s)=\frac{\left[1+(s\tau)^{-1}\right]^{1/2}}{s\left[1+(s\tau)^{-1}\right]^{1/2}+\lambda_n}.
\end{eqnarray}
The solution then becomes
\begin{eqnarray}\label{sol T}
T_{n}(t)&=&\sum_{j=0}^{\infty}(-\lambda_n)^{j}\frac{s^{-j/2-1}}{\left(s+\tau^{-1}\right)^{j/2}}\nonumber\\&=&\sum_{j=0}^{\infty}\left(-\lambda_n\right)^{j}t^{j}E_{1,j+1}^{j/2}\left(-\frac{t}{\tau}\right),
\end{eqnarray}
with $\lambda_n=\frac{\mathcal{D}_{x}}{2\sqrt{\mathcal{D}_{y}\tau}}\left[\frac{(2n+1)\pi}{2L}\right]^{2}$. We note that $T_n(t=0)=1$, since for $t=0$ only the first term with $j=0$ in the sum (\ref{sol T}) survives. Thus, the finite domain solution reads
\begin{eqnarray}\label{sol p1}
p_{1}(x,t)&=&\frac{1}{L}\sum_{n=-\infty}^{\infty}e^{\imath\frac{(2n+1)\pi x}{2L}}\nonumber\\&\times&\sum_{j=0}^{\infty}\left(-\lambda_n\right)^{j}t^{j}E_{1,j+1}^{j/2}\left(-\frac{t}{\tau}\right).
\end{eqnarray}

Accounting relaxation in the finite boundaries, we also find the survival probability $S(t)=\int_{-L}^{L}p_{1}(x,t)\,dx$ which reads
\begin{eqnarray}\label{survival probability}
S(t)=\frac{4}{\pi}\sum_{n=0}^{\infty}\frac{(-1)^n}{2n+1}\sum_{j=0}^{\infty}\left(-\lambda_n\right)^{j}t^{j}E_{1,j+1}^{j/2}\left(-\frac{t}{\tau}\right),
\end{eqnarray}
from where the first passage time PDF is
\begin{eqnarray}\label{survival_probability}
f(t)&=&-\frac{d}{dt}S(t)\nonumber\\&=&\frac{4}{\pi}\sum_{n=0}^{\infty}\frac{(-1)^{n+1}}{2n+1}\sum_{j=0}^{\infty}\left(-\lambda_n\right)^{j}\frac{d}{dt}t^{j}E_{1,j+1}^{j/2}\left(-\frac{t}{\tau}\right).\nonumber\\
\end{eqnarray}
The long time limit can be obtained by asymptotic expansion of the Mittag-Leffler function (\ref{GML_formula}), which results in the following chain of transformations
\begin{eqnarray}\label{survival_probability_asympt}
f(t)&\simeq&\frac{4}{\pi}\sum_{n=0}^{\infty}\frac{(-1)^{n+1}}{2n+1}\sum_{j=0}^{\infty}\left(-\lambda_n\sqrt{\tau}\right)^{j}\frac{t^{j/2-1}}{\Gamma(j/2)}\nonumber\\&=&\frac{4}{\pi}\sum_{n=0}^{\infty}\frac{(-1)^{n+1}}{2n+1}t^{-1}E_{1/2,0}\left(-\lambda_{n}\sqrt{\tau}t^{1/2}\right)\nonumber\\&=&\frac{\pi}{L^2}\frac{\mathcal{D}_x}{2\sqrt{\mathcal{D}_y}}\sum_{n=0}^{\infty}(-1)^n(2n+1)\,t^{-1/2}\nonumber\\&\times&E_{\frac{1}{2},\frac{1}{2}}\left(-\frac{\mathcal{D}_x}{2\sqrt{\mathcal{D}_y}}\frac{(2n+1)^2\pi^2}{4L^2}t^{1/2}\right),
\end{eqnarray}
where we use the relation $E_{\alpha,\beta}(z)=zE_{\alpha,\alpha+\beta}(z)+\frac{1}{\Gamma(\beta)}$ \cite{ml_paper}. From here, we find a power-law decay of the form $f(t)\simeq t^{-3/2}$, i.e.,
\begin{eqnarray}\label{survival_probability_asympt2}
f(t)&\simeq&\frac{8L^2}{\pi^{7/2}}\frac{2\sqrt{\mathcal{D}_y}}{\mathcal{D}_x}\sum_{n=0}^{\infty}\frac{(-1)^n}{(2n+1)^3}\,t^{-3/2}\nonumber\\&=&\frac{L^2}{8\pi^{7/2}}\frac{2\sqrt{\mathcal{D}_y}}{\mathcal{D}_x}\left[\zeta\left(3,\frac{1}{4}\right)-\zeta\left(3,\frac{3}{4}\right)\right]t^{-3/2},\nonumber\\
\end{eqnarray}
where $\zeta(s,a)=\sum_{k=0}^{\infty}\frac{1}{(k+a)^s}$ is the Hurwitz zeta function \cite{saxena_book}. 

The obtained result has a correct limit in the infinite domain $L\rightarrow\infty$. The solution (\ref{sol p1}) in the Laplace space is given by
\begin{eqnarray}\label{sol p1 laplace}
\hat{p}_{1}(x,s)&=&\frac{1}{L}\sum_{n=-\infty}^{\infty}e^{\imath\left[\frac{(2n+1)\pi}{2L}\right]x}\nonumber\\&\times&\frac{1}{s}\frac{s^{1/2}(1+s\tau)^{1/2}}{s^{1/2}(1+s\tau)^{1/2}+\frac{\mathcal{D}_{x}}{\sqrt{2\mathcal{D}_{y}}}\left[\frac{(2n+1)\pi}{2L}\right]^{2}},\nonumber\\
\end{eqnarray}
Taking the limit $L\rightarrow\infty$, i.e., $1/L\rightarrow0$, the summation leads to integration which corresponds to the inverse Fourier transform with $k_{x}=\frac{(2n+1)\pi}{2L}$, from where we obtain an equivalent equation to Eq.~(\ref{p1 FL}) for the reduced PDF for the infinite domain case.

\section{Summary}
We have concerned with a finite velocity of a particle spreading in the framework of a Cattaneo equation in the comb geometry. This issue of anomalous diffusion with finite velocity in comb geometry was already established in numerical studies \cite{flux1,flux2,liu}. We presented exact solutions in the form of the Fox $H$-function series for both initial and boundary value problems. It describes kinetics, which is essentially complicated in comparison with anomalous diffusion described  by comb fractional Fokker-Planck equation, obtained previously for a variety of realizations \cite{iomin2}. Fractional Cattaneo equation~(\ref{p1 eq Pr}) describes a transient process of diffusion-subdiffusion with long time asymptotics of the MSD $\sim t^{1/2}$. For the dendrite transport described by Eq.~(\ref{p1 eq Pr}), the boundary conditions are important. Various boundary value problems for the fractional Fokker-Planck equation have been already discussed in Ref.~\cite{metzler_klafter}. Here we considered a symmetrical boundaries with absorption, which is also important for the relaxation in many applications, including spiny dendrites and also for example in application of multichannel diffusion of hydrogen in solids \cite{yu}.

\acknowledgments
T.S. acknowledges funding from the Deutsche Forschungsgemeinschaft (DFG), project ME 1535/6-1 "Random search processes, L\'evy flights, and random walks on complex networks". A.I. acknowledges the support from the Israel Science Foundation (ISF-931/16).


\begin{thebibliography}{0}

\bibitem{cattaneo1}
Cattaneo C., {\it Atti Semin. Mat. Fis. Univ. Modena Reggio Emilia} {\bf3} (1948) 83.

\bibitem{cattaneo2}
Cattaneo C. R., {\it Comptes Rendus} {\bf247} (1958) 431.

\bibitem{ref1}
Joseph D. D. and Preziosi L., {\it Rev. Mod. Phys.} {\bf61} (1989) 41; {\it Rev. Mod. Phys.} {\bf62} (1990) 375.

\bibitem{masoliver_weiss}
Masoliver J. and Weiss G. H., {\it Eur. J. Phys.} {\bf17} (1996) 190.

\bibitem{weiss}
Weiss G. H., {\it Physica A} {\bf311} (2002) 381.

\bibitem{masoliver_lindenberg}
Masoliver J. and Lindenberg K., {\it Eur. Phys. J. B} {\bf90} (2017) 107; Masoliver J., {\it Phys. Rev. E} {\bf93} (2016) 052107; Masoliver J., {\it Phys. Rev. E} {\bf96} (2017) 022101; Bogu\~{n}\'a M., Porr\`{a} J. M. and Masoliver J., {\it Phys. Rev. E} {\bf58} (1998) 6992.

\bibitem{metzler_compte}
Compte A. and Metzler R., The generalized Cattaneo equation for the description of anomalous transport processes, {\it J. Phys. A: Math. Gen.} {\bf30} (1997) 7277.

\bibitem{colin}
Olivares-Robles M. A. and Garc\'{i}a-Col\'{i}n L. S., {\it J. Non-Equilib. Thermodyn.} {\bf21} (1996) 361.

\bibitem{qi1}
Qi H. and Jiang X., {\it Physica A} {\bf390} (2011) 1876. 

\bibitem{ramirez}
Fernandez-Anaya G., Valdes-Parada F. J. and Alvarez-Ramirez J., {\it Physica A} {\bf390} (2011) 4198.

\bibitem{qi2}
Qi H. -T., Xua H. -Y. and Guo X. -W., {\it Comput. Math. Appl.} {\bf66} (2013) 824; Qi H. and Guo X., {\it Int. J. Heat and Mass Transf.} {\bf76} (2014) 535.


\bibitem{qi4}
Xu H. -Y., Qi H. -T. and Jiang X. -Y., {\it Chin. Phys. B} {\bf22} (2013) 014401.

\bibitem{flux1}
Liu L., Zheng L. and Zhang X., {\it Appl. Math. Model.} {\bf40} (2016) 6663.

\bibitem{zorica}
Cveti\'canin S. M., Zorica D. and Rapai\'c M. R., {\it Nonlin. Dyn.} {\bf88} (2017) 1453.

\bibitem{flux2}
Liu L., Zheng L., Chen Y. and Liu F., {\it J. Stat. Mech.} {\bf2018} (2018) 013208. 

\bibitem{siam}
Ferrillo F., Spigler R. and Concezzi M., {\it SIAM J. Appl. Math.} {\bf78} (2018) 1450.

\bibitem{segal}
Korkotian E. and Segal M., {\it Cell Calcium} {\bf40} (2006) 441.

\bibitem{segal2}
Segal M., {\it Nature Rev. Neurosci.} {\bf6} (2005) 277. 

\bibitem{iomin_csf}
M\'endez V and Iomin A., {\it Chaos Solitons \& Fractals} {\bf53} (2013) 46.

\bibitem{iomin_pre2013}
Iomin A. and M\'endez V., {\it Phys. Rev. E} {\bf88} (2013) 012706.

\bibitem{yuste}
Yuste S. B., Abad E. and Baumgaertner A., {\it Phys. Rev. E} {\bf94} (2016) 012118.

\bibitem{rose}
Rose J., Jin S. -X. and Craig A., {\it Neuron} {\bf61} (2009) 351.

\bibitem{bresslof}
Earnshaw B. A. and Bressloff P. C., {\it J. Comput. Neurosci.} {\bf28} (2010) 77.

\bibitem{santamaria}
Santamaria F., Wils S., De Schutter E. and Augustine G. J., {\it Neuron} {\bf52} (2006) 635.

\bibitem{santamaria2}
Santamaria F., Wils S., De Schutter E. and Augustine G. J., {\it Eur. J. Neurosci.} {\bf34} (2011) 561.

\bibitem{iomin_new}
Iomin A., Zaburdaev V. and Pfohl T., {\it Chaos Solitons \& Fractals} {\bf92} (2016) 115.

\bibitem{saxena_book}
Mathai A. M., Saxena R. K. and Haubold H. J., {\it The $H$-function: Theory and Applications} (New York Dordrecht Heidelberg London, Springer) 2010.

\bibitem{fcaa_delta}
S\"udland N. and Baumann G., {\it Fract. Calc. Appl. Anal.} {\bf7} (2004) 409.

\bibitem{schneider1}
Montroll E. W. and Weiss G. H., {\it J. Math. Phys.} {\bf6} (1965) 167.

\bibitem{hilfer}
Hilfer R., {\it Application of Fractional Calculus in Physics} (Singapore, World Scientiffic Publishing Company) 2000.

\bibitem{mmnp}
Sandev T., Iomin A., Kantz H., Metzler R. and Chechkin A., {\it Math. Model. Natur. Phenom.} (2016) {\bf11} 18.

\bibitem{new}
Lundstrom B. N., Higgs M. H., Spain W. J. and Fairhall A. L., {\it Nature Neuroscience} (2008) {\bf11} 1335.

\bibitem{prabhakar}
Prabhakar T. R., {\it Yokohama Math. J.} {\bf19} (1971) 7.

\bibitem{garra}
Garra R., Gorenflo R., Polito F. and Tomovski Z., {\it Appl. Math. Comput.} {\bf242} (2014) 576.

\bibitem{garrappa}
Garra R. and Garrappa R., {\it Commun. Nonlinear Sci. Numer. Simul.} {\bf56} (2018) 314; Sandev  T., Chechkin A., Korabel N., Kantz H., Sokolov I. M. and Metzler R., {\it Phys. Rev. E} {\bf92} (2015) 042117.

\bibitem{ml_paper}
Haubold H. J., Mathai A. M. and Saxena R.K., {\it J. Appl. Math.} {\bf2011} (2011) 298628.


\bibitem{liu}
Liu L., Zheng L., Chen Y. and Liu F., {\it Commun. Nonlin. Sci. Numer. Simul.} {\bf63} (2018) 135.

\bibitem{iomin2}
Iomin A., Mendez V. and Horsthemke W., \textit{Fractional Dynamics in Comb-like Structures}, (World Scientific, Singapore) 2018.

\bibitem{metzler_klafter}
Metzler R. and Klafter J., {\it Physica A} {\bf278} (2000) 107.

\bibitem{yu}
Andronov D.Yu., Arseniev D. G., Polyanskiy A. M., Polyanskiy V. A. and Yakovlev Yu. A., {\it Int. J. Hydrogen Energy} {\bf42} (2017) 699; Zhang Y., Maeda R., Komaki M. and Nishimura C., {\it J. Membrane Sci.} {\bf269} (2006) 60; Zaika Yu. V. and Bormatova E. P., {\it Tech. Phys.} {\bf55} (2010) 347.

\end{thebibliography}
\end{document}